\documentstyle[11pt,newpasp,epsf,twoside]{article}
\def\edcomment#1{\iffalse\marginpar{\raggedright\sl#1\/}\else\relax\fi}
\reversemarginpar

\begin{document}
\title{Dwarf Nova Oscillations: New Results}

\author{Brian Warner and Patrick A. Woudt} 

\affil{Department of Astronomy, University of Cape Town, Rondebosch 7700, South Africa}

\begin{abstract}
New and archived observations of VW Hyi in outburst show the occasional presence of
optical Dwarf Nova Oscillations (DNOs) over the range of 18 -- 40 s. There is a rapid increase
in period near the end of outburst, at the same time that the EUV falls almost to zero, which
we attribute to propellering. The DNOs return to a shorter period after this phase, but are
very incoherent. The DNOs show some modulation by the Quasi-Periodic Oscillations (QPOs)
that are also occasionally present in the light curve. We interpret the QPOs as a prograde 
travelling wave in the inner disc, which obscures and/or reprocesses radiation from the central
region. The model is applied to observations of OY Car and WZ Sge.
\end{abstract}

\section{Introduction}

The presence of brightness modulations of low amplitude and moderate coherence in Cataclysmic
Variables (CVs) was first detected in 1972 (Warner \& Robinson 1972). Of similar appearance but 
much lower stability than the 71 s modulation in the old nova DQ Her, these oscillations are found 
almost exclusively in dwarf novae during outburst or in nova-like variables. Collectively they 
have become known as Dwarf Nova Oscillations (DNOs).  They are not found in all nova-likes, 
nor in all dwarf novae in outburst, but they have been found in a total of about 15 of the former 
and 4 of the latter (see the list given in Chapter 8 of Warner (1995a)). DNOs are also observed 
in the soft X-ray and EUV regions of some CVs. The shortness of period and the emission 
of high energy radiation indicate that the origin of the DNOs is at, or close to, the surface 
of the white dwarf primary. However, the phase shifts seen during eclipse in the optical 
(e.g.~HT Cas: Patterson 1981; UX UMa: Nather \& Robinson 1974), which extend right through 
eclipse, show that the concave surface of the disc reprocesses high energy radiation from 
a rotating irradiating beam. In the UV the eclipse of the oscillations is of similar 
duration to that of the white dwarf (Knigge et al.~1998), demonstrating directly that the 
source of the energy is close to the primary.

The most striking property of the DNOs is that they show a strong period-luminosity
relationship during the outbursts of dwarf novae: minimum period corresponds to highest 
UV luminosity, i.e.~to the phase when mass transfer on to the primary is at its maximum. 
The range of period during this systematic variation can be a factor of two over the few 
days of an outburst. This is in strong contrast to DQ Her, which has a time scale $\sim$10$^7$ y 
for change of its period, which is understood in terms of accretion of angular momentum onto 
the white dwarf. Therefore, if the two phenomena are similar in underlying physics, only 
a very small fraction of the mass of the white dwarf is participating in angular momentum 
transfer. Paczynski (1978) long ago proposed that this could be understood if the accreting 
matter produced a rapidly spinning equatorial band that spun up during the increasing $\dot{M}$ 
to maximum luminosity and was spun down again in the decline stages of the outburst. 
Katz (1975) estimated that in order to couple the outer regions to the core of a white 
dwarf a magnetic field in excess of 10$^5$ G is necessary. Freely rotating equatorial belts 
would therefore only be expected for field strengths lower than this.

Evidence for equatorial belts is now quite strong -- in VW Hyi HST spectra have been 
interpreted as implying a hot equatorial region rotating at nearly the Kepler velocity 
and lasting for a week or more after outburst (Sion et al.~1996); an equatorial accretion 
belt has also been deduced for U Gem (Cheng et al.~1997). Possible detection of magnetically 
controlled accretion flow in VW Hyi during outburst is given by the existence of an inverse 
P Cygni spectral profile and a truncated disc (Huang et al.~1996).
   
We therefore favour the interpretation of DNOs as magnetically controlled accretion onto 
an equatorial belt -- it becomes an extension of the standard intermediate polar model to 
lower field strengths, and because an essential component is the free slipping of the
equatorial belt we call the model the Low Inertia Magnetic Accretor (LIMA). This model has already 
been developed to some extent by Warner (1995b).

\begin{figure}[h]
\plotfiddle{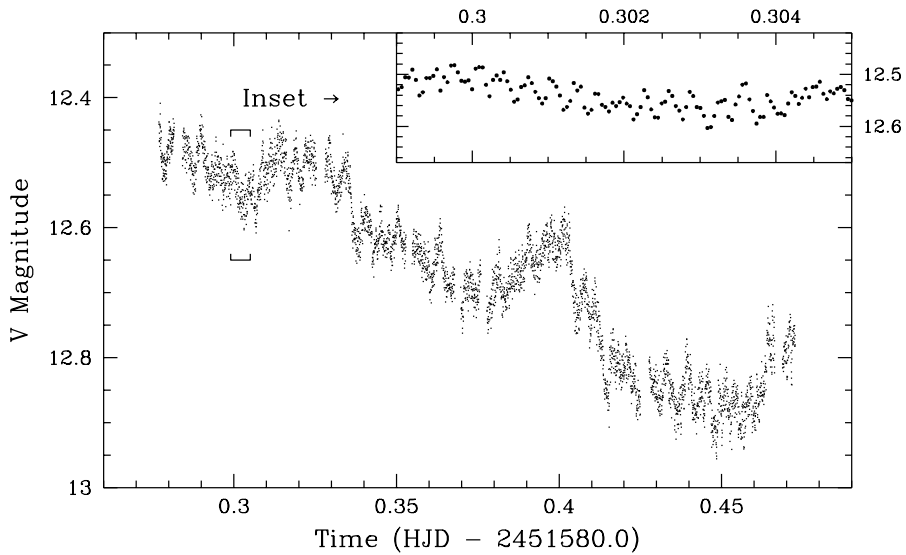}{7.0cm}{0}{70}{70}{-245}{-195}
\caption{The light curve of VW Hyi, taken on 5 February 2000 during the return of VW Hyi to quiescence 
after a normal outburst. An enlarged view of a small section of the light curve shows the DNOs.}
\label{warnerf1}
\end{figure}

\section{New Observations of VW Hyi}

In Fig.~1 we show the light curve of VW Hyi at 3 s time resolution that we obtained 
on 5 February 2000 just as the system was reaching quiescence after a normal outburst. 
The large amplitude modulation is the normal orbital hump. The inset is a small section 
of light curve greatly magnified and shows the presence of $\sim$30 s DNOs that persist 
throughout the run. There is also a $\sim$500 s modulation that we discuss below. 
The light curve is made up almost entirely of these three modulations -- there is very little flickering.

\begin{figure}[h]
\plotfiddle{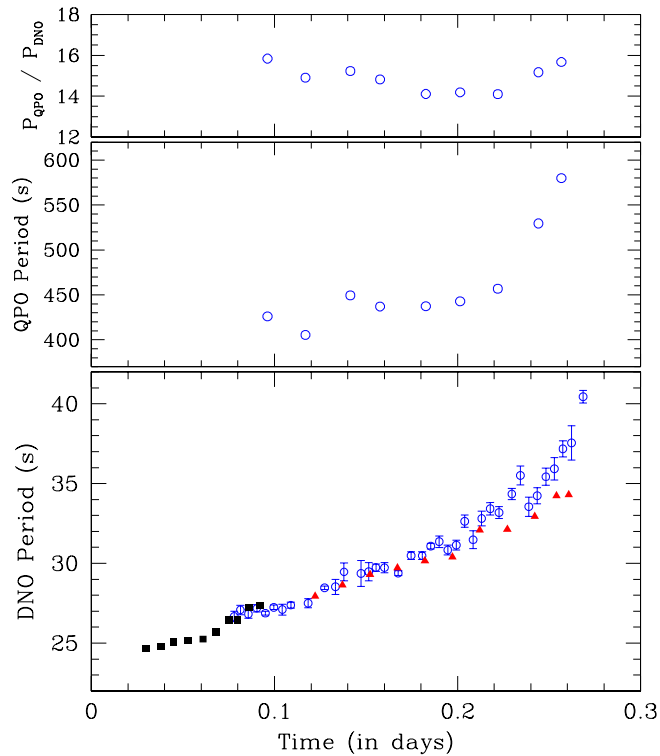}{9.8cm}{0}{55}{55}{-135}{-90}
\caption{Evolution of DNO and QPO periods in three outbursts: February 2000 (circles with error bars),
December 1972 (triangles), and February 2001 (squares). The top panel shows the ratio of periods in
the February 2000 outburst.}
\label{warnerf1}
\end{figure}

\begin{figure}[h]
\plotfiddle{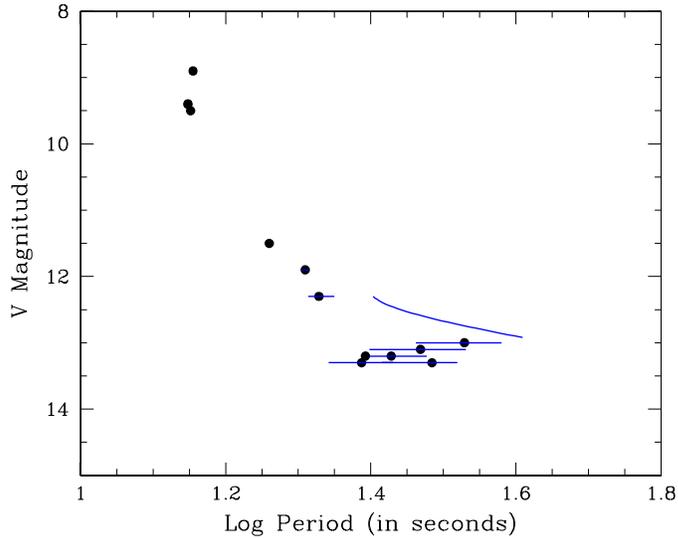}{7.0cm}{0}{50}{50}{-155}{-135}
\caption{DNO periods versus V magnitude of VW Hyi. The curved continuous line is the DNO
evolution seen in Fig.~2. The range of DNO periods in the various runs is shown by horizontal
bars.}
\label{warnerf1}
\end{figure}

The DNOs in this light curve are of relatively large amplitude (most DNOs are only visible 
in a Fourier Transform) and steadily increase their period through the 5.3 h run. 
This is shown in Fig.~2, where we have added the DNO behaviour observed 
at the end of a superoutburst of VW Hyi in December 1972 and at the end of a normal 
outburst in February 2001. We have also reanalysed our archive of VW Hyi high speed photometry 
built up over nearly 30 years, and found some previously overlooked DNOs. In particular, we have 
found a 14.29 s signal near the maximum of the 26 October 1984 superoutburst and a 14.06 s DNO 
near maximum in the 19 December 1982 superoutburst.  These signals, at periods nearly half of 
the shortest periods previously detected in the optical, are pleasingly close to the 14.06 s DNO seen 
in soft X-rays in the November 1983 superoutburst and the periods of 14.2 to 14.4 s in X-rays at 
the peak of the October 1984 outburst (van der Woerd et al.~1987). 
The optical period of 33.9 s seen by Schoembs \& Vogt (1980) near maximum of the October 1978 superoutburst 
is in fact a 14.18 s DNO beating with their integration time of 10 s. The 14.06 s DNO in our data is 
of very high stability, which enables it to be seen easily in the FT despite having an amplitude 
of only 0.0012 mag. It is possible that such short very low amplitude DNOs are commonly present but 
not stable enough to detect with the usual techniques.

These short period DNOs, and others of longer period measured in our archived and recent light curves 
of VW Hyi, lead to the variation of DNO period with brightness in outburst shown in Fig.~3.  
The slope of the variation is at first similar to what has been seen in many dwarf novae in outburst, but 
the rapid increase in period (the continuous curve in Fig.~3, derived from Fig.~2) 
is anomalous and implies a very rapid deceleration (a factor of 2 in angular velocity in about 10 h) of 
the equatorial belt in our LIMA model.

In addition to the DNOs discussed above, CVs commonly show far less coherent oscillations with periods 
an order of magnitude longer -- known as Quasi-Periodic Oscillations (QPOs). The oscillations of range $\sim$0.1 
mag seen in Fig.~1 are QPOs, and are the first in which a systematic evolution of period has been 
seen -- Fig.~2 shows that their mean period increase from $\sim$400 s to $\sim$600 s during the
run. The ratio $P_{QPO}$/$P_{DNO}$ remains roughly constant at $\sim$15 during their evolution in both 
the February 2000 and December 1972 outbursts. One of the features that we have noticed about QPOs is 
that their minima often appear as absorption dips, carrying the minima of
their modulations below the general lower envelope of flickering and other slower modulations -- see Fig.~4.

We have also detected QPOs in VW Hyi in quiescence. This is the first time QPOs in a CV have been found 
at minimum light. As with other CVs (but see WZ Sge below) we do not find any DNOs during quiescence.

\begin{figure}[t]
\plotfiddle{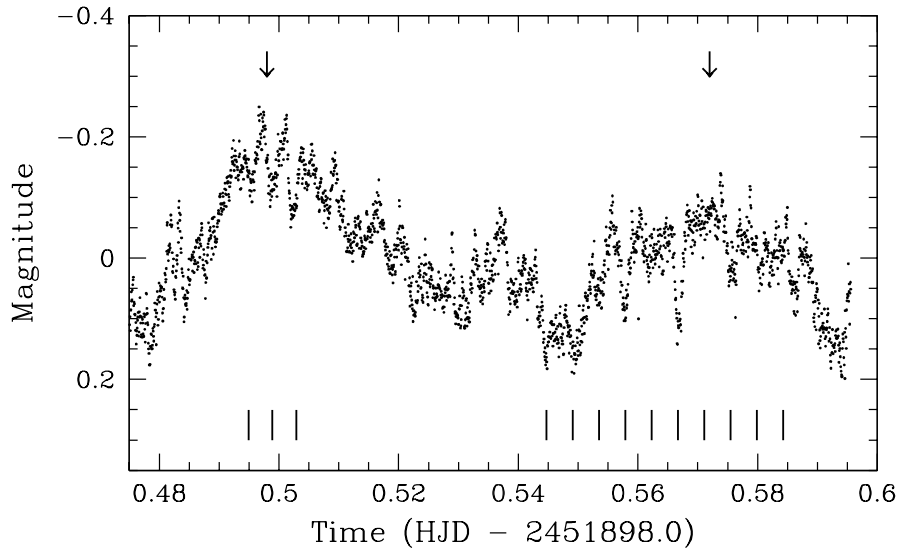}{7.5cm}{0}{70}{70}{-235}{-185}
\caption{Part of a light curve of VW Hyi towards the end of the December 2000 outburst. Prominent
dips are marked by vertical bars, arrows mark times of orbital hump maxima.}
\label{warnerf4}
\end{figure}

\section{Interaction between DNOs and QPOs}
 
In previous studies of VW Hyi (Warner \& Brickhill 1978; Robinson \& Warner 1984) simultaneous DNOs and QPOs
have been observed, and it was noted that at times the QPOs modulate the amplitude of the DNOs.  
In our new observations and analyses we see further occasional examples of this, but there is a 
new phenomenon that has important implications for models of DNOs and QPOs.

In part of a run made during outburst on 3 November 1974 there is a double DNO: DNOs are normally 
monoperiodic (there are a few other cases of double DNOs), but this run has DNOs at 28.77 s and 
31.16 s and a strong QPO at 349 s. This QPO period is the {\it beat period} between the two DNOs. The 
shorter period of the DNOs is purely sinusoidal; the longer period has a first harmonic. 
Average profiles of the three modulations are shown in Fig.~5. The DNO behaviour in VW Hyi is 
exactly similar to what Marsh \& Horne (1998) have found in HST observations of OY Car -- there they see 
two periods near 18 s, separated by 0.22 s, with the longer period having a dominant first harmonic. 
In addition, Steeghs et al.~(2001) found a double DNO in V2051 Oph on the decline from outburst. 
In neither of these latter observations is the light curve sufficiently long to enable certain detection 
of a QPO at the beat period, if one existed.

\begin{figure}[t]
\plotfiddle{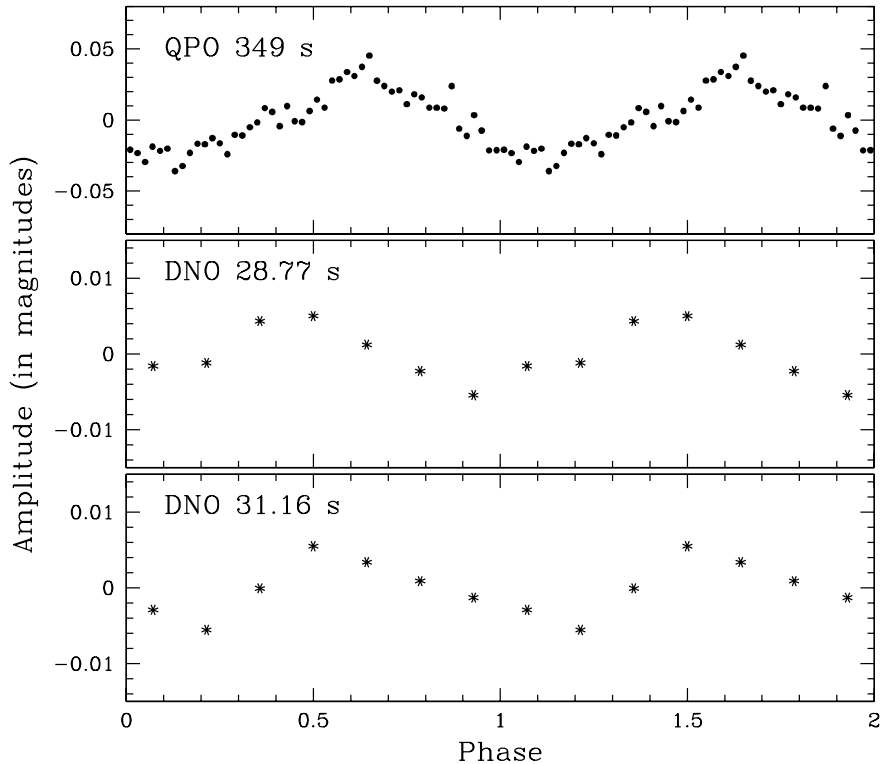}{10.4cm}{0}{70}{70}{-235}{-105}
\caption{Average profiles of the QPO and the two DNOs present in the light curve of VW Hyi
on 3 November 1974.}
\label{warnerf5}
\end{figure}

\section{The Nature of QPOs}

Because of their low coherence QPOs are often more easily seen in the light curve than in its
FT (where they spread over a large range of frequencies). They are rarely seen at high energies, 
but there are a few convincing cases. for example $\sim$500 s at the end of an outburst of VW Hyi 
by Ginga at 2-10 keV (Wheatley et al.~1996), where we see similar time scale optical QPOs.

A popular theoretical explanation of QPOs is that they arise as non-radial oscillations in 
the accretion disc. Perturbation analyses, e.g.~by Carroll et al.~(1985) and Collins, Helfer \& van Horn (2000), 
show that a wide range of periods could exist, but no explanation has been given of what mechanism 
selects the limited range of periods observed in any individual object. On the other hand, Lubow \& Pringle (1993) 
find that an $m=1$ g-mode in the inner disc is the most likely to be excited. This mode corresponds 
to a travelling wave moving upstream in the Keplerian flow, at an angular frequency slightly less 
than that of the flow, so it amounts to a prograde slowly travelling wall.

The theoretical models usually aim to explain QPOs in discs as luminosity variations intrinsic 
to the disc itself (analogous to the kappa or epsilon mechanisms of stellar pulsations). But our 
observations of absorption dips suggests an alternative reason for the brightness modulations -- interception and 
reprocessing of the high energy radiation coming from the accretion-heated white dwarf surface. 
If the travelling wall is very near the white dwarf then it will send additional reprocessed light 
to us when at superior conjunction, and may partially obscure the primary or very innermost part 
of the accretion curtain when at inferior conjunction. Furthermore, the same interception may 
occur for the rotating DNO beam, creating DNO sidebands in analogy with intermediate polars. 
The longer period DNO would be the reprocessed beam for a progradely travelling wall, and 
would be the one likely to depart from sinusoidality (because of irregular structure of the wall), 
just as is observed. Variations in the amplitude of the wall will account for the intermittency 
of production of double DNOs -- indeed, it could happen that in some systems the geometry is 
such that no interception at all of the beam occurs.

The QPOs seen in hard X-rays towards the end of outburst (Wheatley et al.~1996) are probably the 
result of quasi-periodic obscuration of the central source by the wall.
    
For the LIMA model, where the inner disc is truncated by the magnetosphere of the primary, the 
region just outside the inner radius is a likely place for excitation of travelling waves -- through 
the differential winding and reconnection of field lines that couple the inner disc to the primary. 
This has been hinted at in 2D simulations and 3D extrapolations of disc-magnetosphere 
interaction (Uzdensky, Konigl \& Litwin 2001).

\section{The Rapid Deceleration of DNOs}

From the energetics of a VW Hyi outburst we can estimate that the accretion onto the 
primary during a normal outburst is $\sim 2 \times 10^{22}$ g and $\sim$8 times greater in a 
superoutburst. The angular deceleration of the equatorial belt then implies an energy extraction 
of $\sim$3 L$_{\odot}$ for a normal outburst and $\sim$8 times larger for a superoutburst. 
Yet at this time (the end of an outburst) the total luminosity of VW Hyi is only $\sim$0.1 L$_{\odot}$, 
so it is clear that the energy loss from the belt is not radiative. The same situation is seen 
in AE Aqr, where the observed spin-down rate of the primary is equivalent to 4 L$_{\odot}$ 
but this is not radiated (de Jager et al.~1994). In the latter star the explanation is that 
the spin-down torque comes from propellering (e.g., Wynn, King \& Horne 1997), and the energy 
(and angular momentum) extracted from the star goes into gas that is centrifuged away. We propose 
that the same is happening in VW Hyi during the rapid deceleration phase -- but only the equatorial belt 
is involved. In fact, the time scale for spin-down of the primary in AE Aqr is $\sim 1.7 \times 10^7$ y; 
scaling this by the ratio of the mass of the belt in VW Hyi to the mass of the primary gives a 
spin-down time scale of $\sim$10 h (for the same applied magnetic and material torques in 
both cases). This is equal to the time scale that we observe in VW Hyi.
    
It is understandable that VW Hyi gets itself into a propellering state at the end of outburst -- as 
$\dot{M}$ from the disc falls at the end of outburst the radius of the inner edge of the disc moves 
outwards. At the same time the spin-down torque acting on the belt is reducing rapidly, 
and so the belt cannot maintain an equilibrium angular velocity with the inner edge of the disc -- it 
spins more rapidly than the disc and its attached magnetic field centrifuges most of the accreting gas away, 
losing angular velocity until it again reaches equilibrium with the disc. The stability of mass transfer 
in these circumstances has been studied by Spruit \& Taam (1993); the gas is not ejected from 
the system, it is merely moved out to a larger radius in the disc.

There is indirect but strong observational evidence for propellering in VW Hyi. 
Over precisely the $\sim$0.5 day range in which we see the rapid deceleration, Mauche, Mattei \& Bateson (2001) 
find that the EUV flux of VW Hyi drops precipitately almost to zero and recovers afterwards. 
As the EUV flux is a monitor of the accretion flow onto the primary, this shows that gas 
is being prevented from accreting.

\section{Other Systems}

Our proposals for interpreting the rich phenomenology of VW Hyi, of which we have given an 
outline above (more extensive and quantitative results are contained in two papers currently 
in press in MNRAS), enable us to achieve understanding of observations of DNOs and QPOs 
in some other systems, of which we give only two examples here.
 
{\bf OY Car}. As already pointed out above, the double DNOs seen in OY Car (Marsh \& Horne 1998) 
are analogous to what we see in VW Hyi and can be interpreted as a DNO with a QPO sideband. 
Although the beat period of $\sim$1500 s is not visible in the short observing run 
containing the double DNO, a QPO at $\sim$2240 s has been seen in X-Ray observations of OY Car 
about two days after a normal outburst (Ramsay et al.~2001), with the properties of a 
periodically obscuring source, and is probably a longer period evolution from the putative 
$\sim$1500 s QPO during outburst.

{\bf WZ Sge}. WZ Sge is a very long interval SU UMa star which has brightness modulations 
at quiescence with periods of 27.868 s and 28.952 s. It has often been claimed that the shorter 
period has orbital sideband structure like an intermediate polar and that therefore 27.868 s is the 
rotation period of the white dwarf (Patterson 1980; Warner, Tout \& Livio 1996). The measured $v \sin i$ 
of the primary (Cheng et al.~1997) and the recently favoured mass of the primary of $\sim$1.0 M$_{\odot}$ 
give $P_{rot}$ for the primary of 28 $\pm$ 8 s, which is compatible with this proposal.
    
However, the 28.952 s period does not fit the scheme of orbital sidebands. In the light of the double DNOs 
seen in VW Hyi and OY Car, and the occasionally observed first harmonic of the 28.952 s modulation 
(Provencal \& Nather 1997), it seems probable that the longer period of the two oscillations is 
caused by reprocessing off a travelling wall. The beat period of the DNOs (as we shall now call them) 
is 744 s. Such a reprocessing model had already been proposed by Lasota, Kuulkers \& Charles (1999), who 
assumed that the wall must be rotating in the disc at a radius where the Keplerian period is 744 s -- which 
puts it at the outer rim of the accretion disc (but that needs to be revised with the larger primary 
mass now deduced). We propose that the reprocessing wall is a slowly travelling wave close to the 
primary -- if it is at the inner edge of the truncated disc, then the observed DNO period puts 
the inner edge very close to the surface of the primary.

With this model we could hope to see the 744 s beat period in the light curve as the travelling wall 
reprocesses or obscures radiation from the central source. We have found just such a signal 
in a high quality light curve of WZ Sge provided by Dr.~Janet Wood. The principal obscuration 
caused by the wall is the long known dip of variable depth that occurs in the vicinity of 
orbital phase 0.3 (Patterson 1980); the other dips are less easily recognised because some of 
them occur near to eclipse or the deep minimum at orbital phase 0.5.

\section{X-ray Binaries}

The DNO and QPO phenomena in CVs have numerous analogues among the X-ray binaries. In particular,
the ratio $P_{QPO}$/$P_{DNO}$ $\sim$ 15 seen in VW Hyi is the same as the ratio of the two
QPOs commonly seen in X-ray binaries (see Fig.~2 of Psaltis, Belloni \& van der Klis 1999) over
a wide range of frequencies. The CV behaviour is an extrapolation of this to frequencies two
orders of magnitude lower.

\bigskip
\acknowledgements{Our research has been supported by grants from the University of 
Cape Town. We are grateful to Chris Mauche for supplying his EUVE observations of VW Hyi 
and to Janet Wood for allowing us to analyse her light curve of WZ Sge.}

\end{document}